\documentclass[twocolumn,tighten]{aastex631}
\journalinfo{}

\shorttitle{AASTeX v6.3.1 Sample article}
\shortauthors{Selvi et al.}

\graphicspath{{./}{figures/}}

\usepackage{longtable}

\usepackage{array,booktabs}
\usepackage[symbols,nogroupskip,nonumberlist]{glossaries-extra}

\usepackage[normalem]{ulem}
\usepackage{soul}

\usepackage{xcolor}

\glsdisablehyper

\newglossaryentry{B}{ 
	name={\ensuremath{\textbf{B}}},
	description={Magnetic field}}

\newglossaryentry{Bmag}{ 
	name={\ensuremath{B_{0}}},
	description={Magnetic field magnitude}}

\newglossaryentry{yup}{ 
	name={\ensuremath{y_{\textrm{up}}}},
	description={y-coordinate of the upper sheet}}

\newglossaryentry{ylo}{ 
	name={\ensuremath{y_{\textrm{lo}}}},
	description={y-coordinate of the lower sheet}}

\newglossaryentry{d}{ 
	name={\ensuremath{\delta}},
	description={Current sheet thickness}}

\newglossaryentry{basevec}{ 
	name={\ensuremath{\vec{e}}},
	description={Base unit vector}}

\newglossaryentry{4cur}{ 
	name={\ensuremath{ \textbf{J}  }},
	description={Four current density}}

\newglossaryentry{4curcomp}{ 
	name={\ensuremath{ J  }},
	description={Four current density components}}
	
\newglossaryentry{3cur}{ 
	name={\ensuremath{ \textbf{j}  }},
	description={Three current density}}
	
\newglossaryentry{3curcomp}{ 
	name={\ensuremath{ j  }},
	description={Three current density components}}

\newglossaryentry{domainlength}{
	name={\ensuremath{L}},
	description={system length}}

\newglossaryentry{pr}{
	name={\ensuremath{p}},
	description={Pressure, static pressure?}}

\newglossaryentry{ndens}{
	name={\ensuremath{n}},
	description={Number density}}

\newglossaryentry{boltzcon}{
	name={\ensuremath{k_{\mathrm{B}}}},
	description={Boltzmann constante}}
	
\newglossaryentry{temp}{
	name={\ensuremath{T}},
	description={Temperature}}

\newglossaryentry{magnvacperm}{
	name={\ensuremath{\mu}_{0}},
	description={Magnetic vacuum permeability}}
	
\newglossaryentry{thermspread}{
	name={\ensuremath{\Theta}},
	description={Thermal spread}}	

\newglossaryentry{relenthdens}{
	name={\ensuremath{w}},
	description={Relativistic enthaply density}}
	
\newglossaryentry{relspecenth}{
	name={\ensuremath{h}},
	description={Relativistic specific enthaply}}

\newglossaryentry{spol}{
	name={\ensuremath{c}},
	description={Speed of light}}
	
\newglossaryentry{specintenrg}{
	name={\ensuremath{\epsilon}},
	description={Specific internal energy}}
	
\newglossaryentry{flpr}{
	name={\ensuremath{p_{f}}},
	description={Fluid pressure}}
	
\newglossaryentry{mdens}{
	name={\ensuremath{\rho}},
	description={Mass density}}
	
\newglossaryentry{nrspecenth}{
	name={\ensuremath{h_{nr}}},
	description={Non-relativistic specific enthalpy}}
		
\newglossaryentry{relenth}{
	name={\ensuremath{H}},
	description={Relativistic enthalpy}}
		
\newglossaryentry{mass}{
	name={\ensuremath{M}},
	description={Mass}}
		
\newglossaryentry{vol}{
	name={\ensuremath{V}},
	description={Volume}}
		
\newglossaryentry{nrenth}{
	name={\ensuremath{H_{nr}}},
	description={Non-relativistic enthalpy}}
	
\newglossaryentry{polyidx}{
	name={\ensuremath{\hat{\gamma}}},
	description={Polytropic index}}	
	
\newglossaryentry{entralt}{
	name={\ensuremath{S}},
	description={Entropy related quantity}}	
	
\newglossaryentry{magnenrgdens}{
	name={\ensuremath{\mathcal{E}_{\mathrm{B}}}},
	description={Magnetic field energy density}}	

\newglossaryentry{pmass}{
	name={\ensuremath{m}},
	description={Particle mass}}

\newglossaryentry{num}{
	name={\ensuremath{N}},
	description={Number of paricles}}

\newglossaryentry{pfreqel}{
	name={\ensuremath{ \omega_{\mathrm{pe}}  }},
	description={Plasma oscillation frequency}}
	
\newglossaryentry{elch}{
	name={\ensuremath{q}},
	description={Electric charge}}
	
\newglossaryentry{deblength}{
	name={\ensuremath{\lambda_{\mathrm{D}}}},
	description={Debye length}}

\newglossaryentry{lfac}{
	name={\ensuremath{\gamma}},
	description={Lorentz factor}}
	
\newglossaryentry{gfac}{
	name={\ensuremath{g}},
	description={Particle g-factor}}	

\newglossaryentry{magnpr}{
	name={\ensuremath{p_{\mathrm{m}}}},
	description={Magnetic pressure}}
	
\newglossaryentry{3pos}{
	name={\ensuremath{\mathbf{r}}},
	description={3-position (vector)}}
	
\newglossaryentry{3poscomp}{
	name={\ensuremath{r}},
	description={3-position (vector) components}}
	
\newglossaryentry{unitbasevec}{
	name={\ensuremath{\textbf{e}  }},
	description={Unit base vector}}	

\newglossaryentry{3vel}{
	name={\ensuremath{\mathbf{v}}},
	description={3-velocity (vector)}}
	
\newglossaryentry{3velcomp}{
	name={\ensuremath{v}},
	description={3-velocity (vector) components}}

\newglossaryentry{time}{
	name={\ensuremath{t}},
	description={Time}}	

\newglossaryentry{4pos}{
	name={\ensuremath{\mathbf{x}}},
	description={4-position (4-vector)}}
	
\newglossaryentry{4poscomp}{
	name={\ensuremath{x}},
	description={4-position (4-vector) components}}
	
\newglossaryentry{4vel}{
	name={\ensuremath{\mathbf{u}}},
	description={4-velocity (4-vector)}}

\newglossaryentry{4velcomp}{
	name={\ensuremath{u} },
	description={4-velocity (4-vector) components}}

\newglossaryentry{lundquist}{ 
	name={\ensuremath{\mathcal{S}}},
	description={Lundquist number}
}

\newglossaryentry{length}{ 
	name={\ensuremath{L}},
	description={Length}
}

\newglossaryentry{c}{ 
	name={\ensuremath{c}},
	description={Speed of light}
}

\newglossaryentry{eta}{ 
 	name={\ensuremath{\eta}},
 	description={Electrical resistivity}
}

\newglossaryentry{tristan}{
	name={\texttt{Tristan-MP}},
	description={Tristan}
}
 
\newglossaryentry{plmagn}{ 
	name={\ensuremath{\sigma}},
	description={Plasma magnetization}
}

\newglossaryentry{Afour}{  
	name={\ensuremath{\mathsf{A}}},
	description={Electromagnetic four-potential}
}

\newglossaryentry{z}{ 
	name={\ensuremath{z}},
	description={Cartesian z-coordinate}
}

\newglossaryentry{A}{  
	name={\textbf{\ensuremath{A}}},
	description={Electromagnetic vector potential}
}

\newglossaryentry{Acomp}{  
	name={\ensuremath{A}},
	description={Electromagnetic vector potential components}
}

\newglossaryentry{Bcomp}{ 
	name={\ensuremath{B}},
	description={Magnetic field components}
}

\newglossaryentry{E}{ 
	name={ \ensuremath{\textbf{E}} },
	description={Electric field}}

\newglossaryentry{Ecomp}{ 
	name={ \ensuremath{ E  } },
	description={Electric field component}}

\newglossaryentry{y}{ 
	name={\ensuremath{y}},
	description={Cartesian y-coordinate}
}

\newglossaryentry{p}{ 
	name={\ensuremath{p}},
	description={Pressure field}
}

\newglossaryentry{Temp}{ 
	name={\ensuremath{\mathcal{T}}},
	description={Temperature field}
}

\newglossaryentry{rho}{ 
	name={\ensuremath{\rho}},
	description={Fluid density field}
}

\newglossaryentry{adindex}{ 
	name={\ensuremath{\hat{\gamma}}},
	description={Adiabatic index}
}

\newglossaryentry{psi}{ 
	name={\ensuremath{\psi}},
	description={Perturbation amplitude}
}

\newglossaryentry{plbeta}{ 
	name={\ensuremath{\beta}},
	description={Plasma beta}
}

\newglossaryentry{magn}{ 
	name={\ensuremath{\sigma}},
	description={Magnetization}
}

\newglossaryentry{rhob}{
	name={\ensuremath{\rho_{\textrm{b}}}},
	description={Background fluid density field}
}

\newglossaryentry{rhoo}{
	name={\ensuremath{\rho_{\textrm{o}}}},
	description={Fluid over-density field}
}

\newglossaryentry{rhop}{
	name={\ensuremath{\rho_{\textrm{p}}}},
	description={Peak fluid density}
}

\newglossaryentry{xp}{
	name={\ensuremath{x_{\textrm{p}}}},
	description={$x$-coordinate of the perturbation}
}

\newglossaryentry{yp}{
	name={\ensuremath{y_{\textrm{p}}}},
	description={$y$-coordinate of the perturbation}
}

\newglossaryentry{lx}{
	name={\ensuremath{l_{x}}},
	description={Perturbation thickness in the $x$-direction}
}

\newglossaryentry{ly}{
	name={\ensuremath{l_{y}}},
	description={Perturbation thickness in the $y$-direction}
}

\newglossaryentry{x}{
	name={\ensuremath{x}},
	description={Cartesian x-coordinate}
}

\newglossaryentry{t}{
	name={\ensuremath{t}},
	description={Time}
}

\newglossaryentry{pp}{
	name={\ensuremath{p_{\textrm{p}}}},
	description={Peak thermal fluid pressure}
}

\newglossaryentry{force}{
	name={\ensuremath{F}},
	description={Force}
}

\newglossaryentry{std}{
	name={\ensuremath{\xi}},
	description={Standard deviation}
}

\newglossaryentry{enrg}{
	name={\ensuremath{\mathcal{E}}},
	description={Energy}
}

\newglossaryentry{plasmafreq}{
	name={\ensuremath{\omega_{p}}},
	description={Plasma frequency}
}

\newglossaryentry{species}{
	name={\ensuremath{\alpha}},
	description={Species}
}

\newglossaryentry{valfven}{
	name={\ensuremath{v_{\mathrm{A}}}},
	description={Alf\'en velocity}
}

\newglossaryentry{valfven0}{
	name={\ensuremath{v_{\mathrm{A0}  }}},
	description={Initial alf\'en velocity}
}  

\newglossaryentry{pfreq}{
	name={\ensuremath{ \omega_{\mathrm{p}}  }},
	description={Plasma frequency}
}

\newglossaryentry{gfreq}{
	name={\ensuremath{ \omega_{\mathrm{c}}  }},
	description={Larmor gyration cyclotron frequency}
}

\newglossaryentry{skind}{
	name={\ensuremath{ \lambda_{p}  }},
	description={Plasma skin depth}}

\newglossaryentry{gper}{
	name={\ensuremath{ P_{\mathrm{c}}  }},
	description={Particle gyration period}
}

\newglossaryentry{grad}{
	name={\ensuremath{ r_{\mathrm{g}}  }},
	description={Particle Larmor gyration radius}
}

\newglossaryentry{gradhot}{
	name={\ensuremath{ r_{\mathrm{g,\mathrm{hot}}}  }},
	description={Particle Larmor gyration radius of hot particles}
}

\newglossaryentry{pper}{
	name={\ensuremath{ P_{\mathrm{p}}  }},
	description={Plasma oscillation period}
}

\newglossaryentry{f}{
	name={\ensuremath{ f }},
	description={Distribution function}}

\newglossaryentry{emp}{
	name={\ensuremath{ \textbf{T}_{p} }},
	description={Single particle energy-momentum tensor}}

\newglossaryentry{empdens}{
	name={\ensuremath{ \mathcal{T}}_{p}},
	description={Single particle energy-momentum density tensor}}

\newglossaryentry{empt}{
	name={\ensuremath{ \textbf{T}  }},
	description={Total particle energy-momentum tensor}}

\newglossaryentry{emptdens}{
	name={\ensuremath{ \mathcal{T} } },
	description={Total article energy-momentum density tensor}}
	
\newglossaryentry{emptdenscomp}{
	name={\ensuremath{ \mathcal{T}} },
	description={Total article energy-momentum density tensor}}

\newglossaryentry{prt}{
	name={\ensuremath{ \textbf{P}  }},
	description={Total pressure tensor}}
	
\newglossaryentry{prtcomp}{
	name={\ensuremath{ P  }},
	description={Total pressure tensor component}}

\newglossaryentry{elemch}{
	name={\ensuremath{ e  }},
	description={Elementary electric charge}}

\newglossaryentry{recrate}{
	name={\ensuremath{ \mathcal{R} }},
	description={Reconnection rate}}

\newglossaryentry{magnflux}{
	name={\ensuremath{ \Phi }},
	description={Magnetic flux}}

\newglossaryentry{multipl}{
	name={\ensuremath{ \tilde{\lambda} }},
	description={Multiplicity of the pair cascade}}

\newglossaryentry{ndensGJ}{
	name={\ensuremath{ n_{\textrm{GJ}} }},
	description={Goldreic-Julian density}}

\newglossaryentry{angfreqBH}{
	name={\ensuremath{ \Omega_{\textrm{BH}} }},
	description={Angular frequency of black hole}}

\newglossaryentry{radiusgrav}{
	name={\ensuremath{ r_{\textrm{g}} }},
	description={Graviational radius}}

\newglossaryentry{gravconst}{
	name={\ensuremath{ G }},
	description={Gravitational constant}}

\newglossaryentry{masssun}{
	name={\ensuremath{M_{\odot}}},
	description={Solar mass}}
 
\newglossaryentry{press}{ 
	name={\ensuremath{p}},
	description={Pressure}
}

\newglossaryentry{boltzconst}{ 
	name={\ensuremath{k_{\textrm{B}}}},
	description={Boltzmann constant}
}

\newglossaryentry{radconst}{ 
	name={\ensuremath{a}},
	description={Radiation constant}
}

\newglossaryentry{elemcswidth}{ 
	name={\ensuremath{\delta_{c}}},
	description={Elementary current sheet width}
}

\newglossaryentry{etaeff}{ 
	name={\ensuremath{\eta_{\textrm{eff}} }},
	description={Collisionless effective resistivity}
}

\begin{document}


 
\title{Effective resistivity in relativistic collisionless reconnection}

\author[0000-0001-9508-1234]{S. Selvi}
\affiliation{Anton Pannekoek Institute, Science Park 904, 1098 XH, Amsterdam, The Netherlands}

\author[0000-0002-4584-2557]{O. Porth}
\affiliation{Anton Pannekoek Institute, Science Park 904, 1098 XH, Amsterdam, The Netherlands}

\author[0000-0002-7301-3908]{B. Ripperda}
\affiliation{School of Natural Sciences, Institute for Advanced Study, 1 Einstein Drive, Princeton, NJ 08540, USA}
\affiliation{NASA Hubble Fellowship Program, Einstein Fellow}
\affiliation{Princeton University, Department of Astrophysical Sciences, 4 Ivy Ln, Princeton, NJ 08544, USA}
\affiliation{Flatiron Institute, Center for Computational Astrophysics, 162 fifth avenue, 10010, New York, NY, USA}

\author[0000-0002-7526-8154]{F. Bacchini}
\affiliation{Centre for mathematical Plasma Astrophysics, Department of Mathematics, KU Leuven, Celestijnenlaan 200B, B-3001 Leuven, Belgium}
\affiliation{Royal Belgian Institute for Space Aeronomy, Solar-Terrestrial Centre of Excellence, Ringlaan 3, 1180 Uccle, Belgium}

\author[0000-0002-1227-2754]{L. Sironi}
\affiliation{Columbia University, Department of Astronomy, 550 west 120th, New York, NY, 10027 USA}

\author[0000-0003-3544-2733]{R. Keppens}
\affiliation{Centre for mathematical Plasma Astrophysics, Department of Mathematics, KU Leuven, Celestijnenlaan 200B, B-3001 Leuven, Belgium}

\begin{abstract}
Magnetic reconnection can power spectacular high-energy astrophysical phenomena by producing non-thermal energy distributions in highly magnetized regions around compact objects. 
By means of two-dimensional fully kinetic particle-in-cell (PIC) simulations we investigate relativistic collisionless plasmoid-mediated reconnection in magnetically dominated pair plasmas with and without guide field. 
In X-points, where diverging flows result in a non-diagonal thermal pressure tensor, a finite residence time for particles gives rise to a localized collisionless effective resistivity. 
Here, for the first time for relativistic reconnection in a fully developed plasmoid chain we identify the mechanisms driving the non-ideal electric field using a full Ohm's law by means of a statistical analysis based on our PIC simulations.
We show that the non-ideal electric field is predominantly driven by gradients of nongyrotropic thermal pressures.
We propose a kinetic physics motivated non-uniform effective resistivity model, which is negligible on global scales and becomes significant only locally in X-points, that captures the properties of collisionless reconnection with the aim of mimicking its essentials in non-ideal magnetohydrodynamic descriptions.
This effective resistivity model provides a viable opportunity to design physically grounded global models for reconnection-powered high-energy emission.
\end{abstract}

\keywords{
\href{http://astrothesaurus.org/uat/739}{High energy astrophysics (739)}; 
\href{http://astrothesaurus.org/uat/1261}{Plasma astrophysics (1261)}; 
\href{http://astrothesaurus.org/uat/288}{Compact objects (288)}; 
\href{http://astrothesaurus.org/uat/994}{Magnetic fields (994)}; 
\href{http://astrothesaurus.org/uat/1393}{Relativity (1393)}; 
\href{http://astrothesaurus.org/uat/1964}{Magnetohydrodynamics (1964)} 
}

\section{Introduction} 
\label{sec:introduction} 
In magnetospheres, jets, and accretion disk coronae of black holes and neutron stars, relativistic magnetic reconnection is widely conjectured as the mechanism powering many spectacular high-energy phenomena such as  
flares from black hole magnetospheres
\citep{Nathanail2020,Ripperda2020,Chashkina2021,Nathanail2022,Ripperda2022}, 
black hole accretion disk coronae \citep{Sironi2020,Sridhar2021,Sridhar2022},
neutron star binary merger precursors \citep{Most2020},
neutron star--black hole post-merger emission \citep{Lyutikov2011,Bransgrove2021}, 
fast radio bursts \citep{Lyubarsky2020,Lyutikov2020,Mahlmann2022,Most2022b},
coherent radio and gamma-ray pulsar magnetospheric emission \citep{Lyubarsky2018,Philippov2018,Philippov2019,Hu2021}, 
giant magnetar flares \citep{Parfrey2012},
gamma-ray flares from pulsar wind nebulae \citep[]{Cerutti2012,Lyubarsky2012}
and flares from blazar jets \citep{Giannios2013,Petropoulou2016}.

The highly magnetized predominantly pair plasma in these environments acts as a reservoir of magnetic energy that can be converted by magnetic reconnection into kinetic and thermal energy and radiation, powering the high-energy emission. 
The majority of these systems is collisionless, meaning that the mean free path of electrons is (much) larger than the system size, such that particles can accelerate to high energies and form non-thermal energy distributions in reconnecting current sheets. 
Reconnection occurring in the collisionless regime requires a kinetic description. 
However, the typical scale separation between plasma scales and the global scales of astrophysical systems is extremely large, such that kinetic descriptions, e.g.\ widely employed particle-in-cell (PIC) methods, are unaffordable to use at realistic scales. 
Fluid-type descriptions like magnetohydrodynamics (MHD) are typically more suitable for modeling global systems, but are by construction collisional and, therefore, unable to capture collisionless reconnection. 
The desire to accurately model reconnection-powered phenomena and 
their high-energy emission in global systems asks for the incorporation of the properties of collisionless reconnection in an MHD description.
In the standard relativistic collisional one-fluid description an explicit finite resistivity can serve as a macroscopic proxy \citep{DelZanna2016,Ripperda2019a,Ripperda2019b} for breaking Alfv\'en's frozen-in condition, thereby enabling reconnection.  
Up until now, a physically motivated model for the resistivity in the relativistic reconnection regime is lacking, where the inability of MHD to fully capture the properties of collisionless reconnection is exposed by, for example, the discrepancy of a factor $\sim 10$ between collisional and collisionless reconnection rates in the plasmoid regime \citep{Uzdensky2010,ComissoBhattacharjee2016}.
This difference may affect the typical timescales of reconnection-powered high-energy emission \citep{Bransgrove2021}.

Our goal is to propose a kinetic physics motivated model for the resistivity which will pave the way for modelling of global systems, while taking the self-consistent properties of collisionless reconnection into account.
To be able to propose such an effective resistivity model, in this work we investigate relativistic collisionless pair plasma reconnection by means of two-dimensional (2D) fully kinetic PIC simulations.
Here, for a fully developed plasmoid chain we identify the reconnection mechanisms providing the non-ideal electric field using a full Ohm's law.
In contrast to previous studies, we analyze self-consistently formed secondary X-points in a fully developed plasmoid chain, instead of primary X-points dominated by initial conditions.
We present a statistical analysis of all (secondary) X-points in the full plasmoid chain during its entire statistical steady state.
The effective resistivity model proposed in this work provides a viable opportunity to mimic the essential properties of relativistic collisionless reconnection in a pair plasma in non-ideal magnetohydrodynamic simulations and design physically grounded global models for reconnection-powered high-energy emission from magnetized, (near-)collisionless systems.

\section{Numerical methods and setup}
\label{sec:Setup}
Relativistic reconnection is studied in a 2D doubly periodic double current sheet system modelling an idealized magnetic reconnection region. 
Herein, two regions of anti-parallel magnetic field lines \citep[in MHD setting adopted from][]{Keppens2013} are embedded in a pair plasma initialized in an equilibrium configuration. 
The magnetic field strength $\gls{Bmag}$ is parameterized by the plasma magnetization 
$\gls{magn} 
= 
\gls{Bmag}^{2} 
\mathbin{/} 
4 \pi \gls{ndens}_{b} \gls{pmass} \gls{spol}^{2} $ 
of the cold background plasma, with $\gls{ndens}_{b}$ the total background plasma number density and $\gls{pmass}$ the electron (and positron) mass.
We focus on the relativistic regime with $\gls{magn} = 10$ giving an initial upstream Alfv\'en speed $\gls{valfven0} = \gls{spol} \sqrt{\gls{magn} \mathbin{/} (1 + \gls{magn})} \approx 0.95 \gls{spol}$.
For the kinetic simulations we use the relativistic multi-species PIC code \texttt{Tristan-MP v2} \citep{Hakobyan2020}. 
We performed three simulations, one without guide field and two with guide field strength $0.25\gls{Bcomp}_{0}$ and $0.5 \gls{Bcomp}_{0}$ up to a final time $\gls{time} = 2 \left[ \gls{domainlength}_{x} / c \right]$.
For guide field strengths $\left[0,0.25\right]\gls{Bcomp}_{0}$ we use a square 2D simulation domain with size $\left[\gls{domainlength}_{x}, \gls{domainlength}_{y}\right] = \left[1000, 1000\right]  \gls{spol} / \gls{pfreq} $, with $ \gls{spol} / \gls{pfreq} $ the skin depth and $\gls{pfreq}$ the plasma frequency, and initialize $160$ particles per cell in total.
For the simulation with guide field strength $0.5\gls{Bcomp}_{0}$ we use a domain size $\left[\gls{domainlength}_{x}, \gls{domainlength}_{y}\right] = \left[1000, 2000\right] c / \gls{pfreq}$ and initialize $40$ particles per cell in total. 
In all simulations, we employ an initial current sheet width $\gls{d}= 5  \gls{spol} / \gls{pfreq}$ and resolve the upstream skin depth with $10$ grid cells.
Throughout this work all quantities are defined with respect to the simulation rest frame and cgs units are used.

\section{Results}
\label{sec:results}
Magnetic reconnection is initiated by an imposed small local magnetic field perturbation at the center of each current sheet triggering a fast growing tearing mode.
Reconnection fronts, propagating along the current sheet at approximately the Alfv\'en speed $\gls{valfven}\approx \gls{spol}$, sweep up the hot particles that provided the initial pressure support.
Inside the large periodic simulation domain we restrict our analysis to a smaller central region of interest, with length $\gls{domainlength}_{x} / 3$, in which the properties of the resulting reconnecting current sheet in between the reconnection fronts are not subject to initialization anymore after the two reconnection fronts have left this region. After that time, the plasma inflow velocity into the region and magnetic energy decay rate remain relatively constant, thereby establishing a statistical steady state in the analysis domain, which effectively has open boundaries within the larger simulation domain.

\begin{figure*}[ht!]
\centering
\resizebox{\textwidth}{!}{ 
\plotone{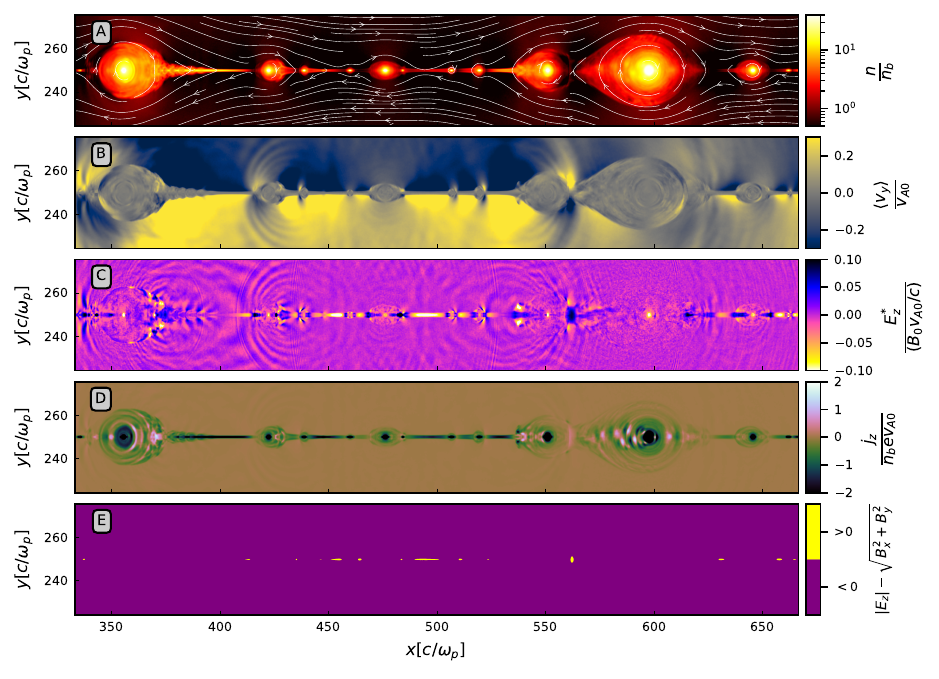}
}
\caption{
Spatial distribution of relevant quantities in the central region with length $\gls{domainlength}_{x} / 3$ at $\gls{time} = 0.7 [ \gls{domainlength}_{x} / \gls{spol} ]$ during the statistical steady state of the simulation without guide field.
From top to bottom panel: 
(panel A) 
the particle number density $\gls{ndens}$ in units of the initial upstream number density $\gls{ndens}_{b}$ showing a hierarchical plasmoid chain as overdense regions;
(panel B)
the mean particle inflow velocity $\langle \gls{3velcomp}_{y} \rangle$ in units of the initial upstream Alfv\'en velocity $\gls{valfven0}$ showing plasma (continuously) being advected into the central region by the reconnection process; 
(panel C)
the out-of-plane non-ideal electric field $\gls{Ecomp}_{z}^{*}$ in units of $\gls{Bcomp}_{0} \gls{valfven0} / \gls{spol}$ showing regions where accelerating reconnection electric fields are created in X-points and magnetic energy is dissipated;
(panel D) 
the out-of-plane current density $\gls{3curcomp}_{z}$ in units of $\gls{ndens}_{b}  \gls{elemch}  \gls{valfven0}$;
(panel E) 
reconnection regions (in yellow) defined as $| \gls{Ecomp}_{z} | > (\gls{Bcomp}_{x}^{2} + \gls{Bcomp}_{y}^{2})^{1/2}$ identifying the non-ideal regions of interest in X-points where reconnection occurs.
}
\label{fig:2Ddomainpanels}
\end{figure*}

Figure \ref{fig:2Ddomainpanels} shows the spatial distribution of various quantities of interest, discussed next, in the central analysis region during this statistical steady state at $\gls{time} = 0.7 [\gls{domainlength}_{x} / \gls{spol}]$ for the simulation without guide field. 
As the current sheet develops a hierarchical plasmoid chain (panel A), plasma is continuously advected into the central region (panel B) by the reconnection process and pushed out along the sheet. 
The non-ideal electric field $\gls{Ecomp}_{z}^{*}$ is the non-vanishing electric field in the frame co-moving with the mean particle velocity (i.e.\ the co-moving frame).
In magnetic X-points, where the magnetic field reconnects (i.e.\ changes topology), an out-of-plane non-ideal electric field $\gls{Ecomp}_{z}^{*}$ (panel C) is created and magnetic energy is dissipated.
Furthermore, an out-of-plane current density $\gls{3curcomp}_{z}$ (panel D) is created in the reconnection layer. 

We investigate the reconnection process in non-ideal reconnection regions (from here on: reconnection regions), defined as $| \gls{Ecomp}_{z} | > (\gls{Bcomp}_{x}^{2} + \gls{Bcomp}_{y}^{2})^{1/2}$ (panel E)  with $\gls{Ecomp}_{z}$ the total out-of-plane electric field, where for the current sheet of Figure \ref{fig:2Ddomainpanels}, $\gls{Ecomp}_{z} < 0 $ is expected for X-points of the main current sheet and $\gls{Ecomp}_{z} > 0 $ in between merging plasmoids \citep{Sironi2022}. 
Comparison of panels C and E shows that this criterion captures well regions where the non-ideal electric field is strong.

Also with guide field the non-ideal regions are well-captured by this definition.
In reconnection regions like those located in the main current sheet and between plasmoid mergers (e.g.\ Figure \ref{fig:2Ddomainpanels} at $x = 565 [\gls{spol} / \gls{pfreq} ]$), usually $\gls{Ecomp}_{z}^{*}$ and $\gls{3curcomp}_{z}$ are aligned. 
We ignore, at any time on average $\sim 1-2$ regions in which they are anti-aligned.
Without guide field we observe $\sim 10$ X-points in the analysis region simultaneously (Figure \ref{fig:2Ddomainpanels}, panel E), while the guide field cases on average generate fewer X-points.

In order to capture the properties of collisionless reconnection it is essential to describe the non-ideal electric fields in the reconnection regions (i.e.\ the reconnection electric field).
Therefore, in the following we identify the mechanisms providing the reconnection electric field and determine their individual contributions using a full Ohm's law for a pair plasma based on our PIC simulations.

\subsection{Reconnection in a single X-point}
\label{sec:singlexpoint}
For a full Ohm's law derived from first principles, by combining the momentum equations for positrons and electrons \citep[]{HesseZenitani2007} and using that masses are the same and charges are opposite in a pair plasma, the out-of-plane $z$-component is,
\begin{widetext}
\begin{eqnarray} 
\gls{Ecomp}_{z} 
&&=  
\underbrace{
-
\left( 
\frac{1}{\gls{spol}} \langle \gls{3vel}_{t} \rangle \times \gls{B} 
\right)_{z}  
}_{\textcolor{black}{\textrm{z-ideal}}} 
\underbrace{
+
\frac{\gls{pmass}}{
\gls{ndens}_{t} \gls{elemch}}
\left( 
\gls{ndens}_{p}  \partial_{\gls{time}}  \langle \gls{4velcomp}_{pz} \rangle 
-
\gls{ndens}_{e}  \partial_{\gls{time}}  \langle \gls{4velcomp}_{ez} \rangle  
\right) 
}_{\textcolor{cyan}{\textrm{z-temporal}}} 
\nonumber \\ && 
\underbrace{
+\frac{\gls{pmass}}{
\gls{ndens}_{t} \gls{elemch}}
\left(
\gls{ndens}_{p} \langle \gls{3velcomp}_{px} \rangle  \partial_{x}  \langle \gls{4velcomp}_{pz} \rangle 
-
\gls{ndens}_{e} \langle \gls{3velcomp}_{ex} \rangle  \partial_{x}  \langle \gls{4velcomp}_{ez} \rangle   
+
\gls{ndens}_{p} \langle \gls{3velcomp}_{py} \rangle \partial_{y}  \langle \gls{4velcomp}_{pz} \rangle 
- 
\gls{ndens}_{e} \langle \gls{3velcomp}_{ey} \rangle \partial_{y}  \langle \gls{4velcomp}_{ez} \rangle 
\right)
}_{\textcolor{teal}{\textrm{z-convective}}} 
\nonumber \\&&
\underbrace{
\underbrace{
+
\frac{1}{
\gls{ndens}_{t} \gls{elemch}}
\left( 
 \partial_{x} \gls{emptdenscomp}_{pxz} 
- 
 \partial_{x} \gls{emptdenscomp}_{exz} 
\right)
}_{\textcolor{brown}{\textrm{xz-total}}}  
\underbrace{
-
\frac{\gls{pmass}}{
\gls{ndens}_{t} \gls{elemch}}
\left( 
\partial_{x} \left(  \gls{ndens}_{p} \langle \gls{3velcomp}_{px} \rangle \langle \gls{4velcomp}_{pz} \rangle \right)  
-
\partial_{x} \left(  \gls{ndens}_{e} \langle \gls{3velcomp}_{ex} \rangle \langle \gls{4velcomp}_{ez} \rangle \right)
\right)
}_{\textcolor{orange}{\textrm{xz-ram}}} 
}_{\textcolor{magenta}{\textrm{xz-thermal}}}
\nonumber \\&&
\underbrace{
\underbrace{
+
\frac{1}{
\gls{ndens}_{t} \gls{elemch}}
\left( 
\partial_{y} \gls{emptdenscomp}_{pyz} 
- 
\partial_{y} \gls{emptdenscomp}_{eyz} 
\right)
}_{\textcolor{olive}{\textrm{yz-total}}}  
\underbrace{
-
\frac{\gls{pmass}}{
\gls{ndens}_{t} \gls{elemch}}
\left( 
\partial_{y} \left( \gls{ndens}_{p} \langle \gls{3velcomp}_{py} \rangle \langle \gls{4velcomp}_{pz} \rangle \right)
- 
\partial_{y} \left(  \gls{ndens}_{e} \langle \gls{3velcomp}_{ey} \rangle \langle \gls{4velcomp}_{ez} \rangle \right) 
\right)
}_{\textcolor{red}{\textrm{yz-ram}}}
}_{\textcolor{violet}{\textrm{yz-thermal}}}.
\label{eq:ohmslawz}
\end{eqnarray}
\end{widetext}
Here,
$
\langle \gls{3vel}_{t} \rangle 
=  
\left( 
\gls{ndens}_{p} \langle \gls{3vel}_{p} \rangle 
+
\gls{ndens}_{e} \langle \gls{3vel}_{e} \rangle 
\right)
/ 
\gls{ndens}_{t} 
$ is the mean total particle 3-velocity, 
$\langle \gls{3vel}_{p} \rangle 
= 
\langle \gls{4vel}_{p} /  \gls{lfac}_{p} \rangle$ 
and 
$\langle \gls{3vel}_{e} \rangle 
= 
\langle \gls{4vel}_{e} /  \gls{lfac}_{e} \rangle$ 
the mean positron and electron 3-velocity,
$\gls{ndens}_{t} = \gls{ndens}_{p} + \gls{ndens}_{e}$ the total number density,
$\gls{ndens}_{p} $ and $ \gls{ndens}_{e} $ the positron and electron number density,
$\langle \gls{4vel}_{p} \rangle $ and $\langle \gls{4vel}_{e} \rangle$ the spatial part of the mean positron and electron 4-velocity,
$\gls{elemch}$ the elementary electric charge,  
$\gls{emptdens}_{\alpha} = \int \left( \gls{pmass}_{\alpha}  \gls{4vel}_\alpha \gls{4vel}_\alpha / \gls{lfac}_\alpha \right) \gls{f}_{\alpha} d^{3} \gls{4vel}$ the spatial part of the symmetric per-species total energy-momentum density tensor,
$\gls{4vel}_{\alpha}$ the spatial part of the particle 4-velocity,
$\gls{lfac}_{\alpha}$ the particle Lorentz factor, and 
$\gls{f}_{\alpha}$ the per-species particle distribution function.

For later reference, terms in equation (\ref{eq:ohmslawz}) are labeled with names that correspond to their physical interpretations and given colors for convenience.
The left-hand side represents the total electric field $\gls{Ecomp}_{z}$ in the $z$-direction. 
The first term labeled $z$-\textit{ideal}) on the right-hand side is the ideal electric field, which vanishes in the co-moving frame.
The remaining terms make up the non-ideal electric field. 
The second term ($z$-\textit{temporal}) arises from temporal changes of the species mean particle momenta,
the third term from convective changes of the species mean particle momenta ($z$-\textit{convective}), 
the fourth and sixth term from the divergence of the total energy-momentum density tensor ($xz$-\textit{total} and $yz$-\textit{total})
and 
the fifth and seventh term from the divergence of the ram pressure tensor ($xz$-\textit{ram} and $yz$-\textit{ram}). 
The total energy-momentum density tensor and ram pressure tensor can be combined into a non-Lorentz-invariant and asymmetric tensor
$
\gls{prt}_{\alpha} 
= 
\gls{emptdens}_{\alpha} - \gls{pmass}_{\alpha} \gls{ndens}_{\alpha} \langle \gls{3vel}_{\alpha} \rangle \langle \gls{4vel}_{\alpha} \rangle 
$, 
which is named throughout this work as the thermal pressure tensor.
This tensor shows some analogies to the thermal pressure tensor in the non-relativistic case, but the actual thermal pressure tensor is only well-defined in the co-moving frame (a discussion on this tensor is given by \citealt{HesseZenitani2007}).
The combination of the fourth with the fifth term and 
of the sixth with the seventh term then respectively results in $xz$-\textit{thermal} and $yz$-\textit{thermal} arising from the divergence of the thermal pressure tensor.

\begin{figure*}[ht!]
\centering
\resizebox{\textwidth}{!}{ 
\plotone{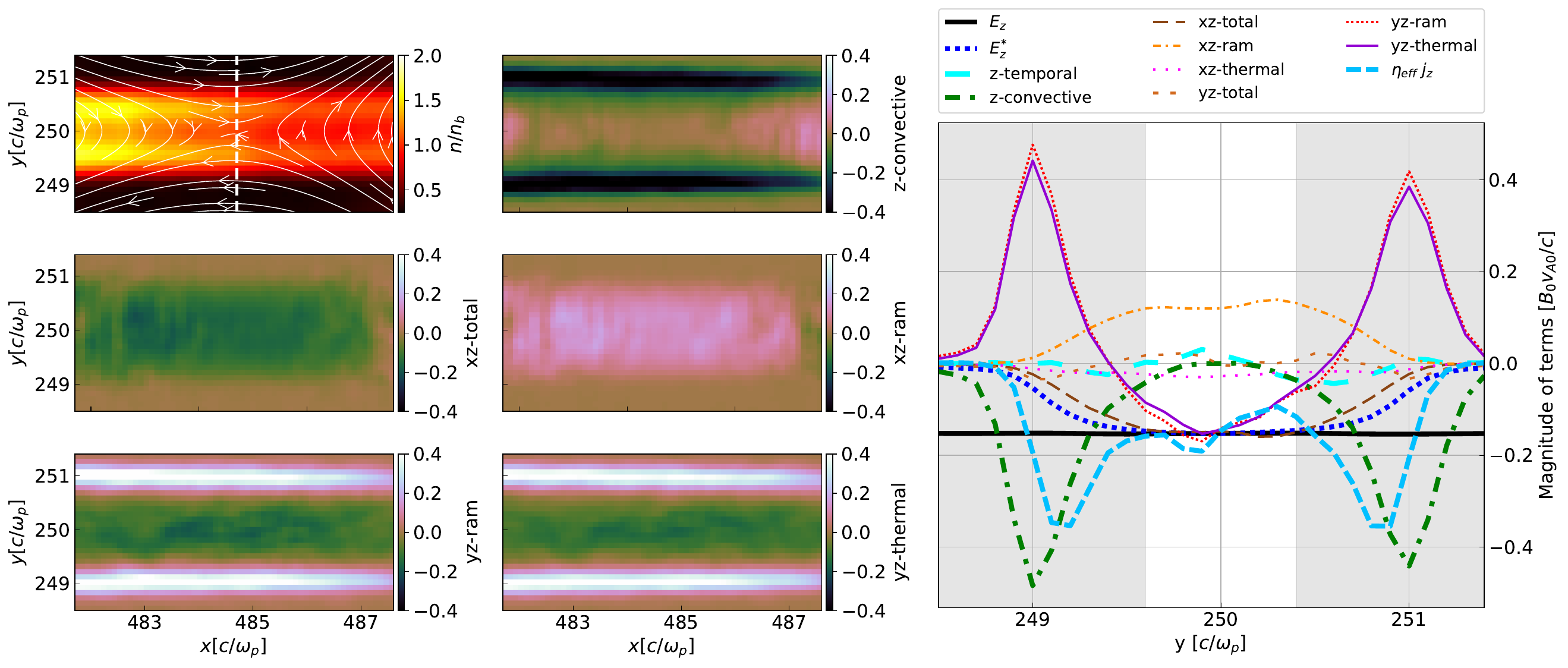}
}
\caption{
The spatial distributions of the density and relevant Ohm’s law terms in a typical X-point (indicated by the magnetic field lines as arrowed white lines in the top left panel) in the plasmoid chain without guide field at $t=0.5 \left[\gls{domainlength}_{x} / \gls{spol} \right]$ in the $6$ left and middle panels. 
In the right panel, (colored) profiles across the current sheet (along the white dashed line in top left panel) with the reconnection region indicated by the non-shaded region. 
Near an X-point the reconnection layer has a substructure that is formed by the individual profiles of contributions of Ohm's law terms.
Inside the reconnection region the reconnection electric field is primarily supplied by gradients of momentum fluxes.
The electric field $\gls{etaeff} \gls{3curcomp}_{z}$ is an adequate approximation to the non-ideal electric field $\gls{Ecomp}_{z}^{*}$ inside the reconnection region}.
\label{fig:2DdomainsAndCuts}
\end{figure*}

We first investigate the reconnection process in a single typical secondary X-point in the plasmoid chain before applying a statistical analysis to the full plasmoid chain.
Figure \ref{fig:2DdomainsAndCuts} shows the spatial distributions of the density and relevant Ohm's law terms (left and middle panels) around an X-point in the plasmoid chain without guide field at $\gls{time} = 0.5 [ \gls{domainlength}_{x} / \gls{spol} ]$ and $x = 485 [\gls{spol} / \gls{pfreq} ]$. 
Additionally, it shows (colored) profiles (right panel) of Ohm's law terms across the overdense current sheet (along the white dashed line in the top panel in the left column) through the X-point (indicated by the magnetic field lines as arrowed white lines). 
The reconnection region (i.e.\ $| \gls{Ecomp}_{z} | > (\gls{Bcomp}_{x}^{2} + \gls{Bcomp}_{y}^{2})^{1/2}$) is indicated as the non-shaded region. 
All Ohm's law terms in Figure \ref{fig:2DdomainsAndCuts} are in units of $\gls{Bcomp}_{0} \gls{valfven0} / \gls{spol}$. 

In the upstream, as expected, all non-ideal terms, i.e. the non-ideal electric field $\gls{Ecomp}_{z}^{*}$ (blue line in the right panel), vanish.
Particles move from upstream regions towards the current sheet from both sides carrying mainly $y$-momentum.
The $z$-temporal term (cyan line in the right panel) has relatively small fluctuations around zero in the entire region.

Inside the reconnection region $\langle \gls{3velcomp}_{ \alpha z} \rangle $ peaks, while $\langle \gls{3velcomp}_{ \alpha y} \rangle $ decreases and changes sign at the center, where plasma flow stagnates coming from both sides.
Furthermore, $\langle \gls{3velcomp}_{ \alpha z} \rangle$ only shows small variations along $x$.
This leads to $\gls{ndens}_{\alpha} \langle \gls{3velcomp}_{ \alpha y} \rangle \partial_{y} \langle \gls{4velcomp}_{ \alpha z} \rangle $ being the dominant term among the $z$-convective terms (top panel in the middle column and green line in the right panel) with peaks just outside while vanishing inside the reconnection region. 

Next, we consider terms, originating from the divergence of pressure tensors, that include gradients of momentum fluxes (i.e.\ gradients of shear stresses).  
Throughout this work we indicate momentum fluxes, being the transfer of the $k$-th component of linear momentum into the $l$-direction, as $kl$-momentum flux. 
Considering the dynamics around an X-point described above, ram $yz$-momentum flux peaks outside and decreases inside the reconnection region, where it changes sign at the center. 
This means that $yz$-ram (i.e.\ the $y$-gradient of ram $yz$-momentum flux; bottom panel in the left column and red line in the right panel), peaks just outside and with opposite sign peaks again inside the reconnection region.

Particles from the upstream pass the magnetic field separatrices and are pushed downstream away from the X-point on both sides along the current sheet out of the reconnection region.
This leads to an increase of their $x$-momentum and, therefore, their $xz$-momentum flux that changes sign on the two sides of the X-point, while further downstream it decreases as particles are not accelerated along $x$ and $z$ anymore.
This means that $xz$-ram (i.e.\ the $x$-gradient of $xz$-momentum flux) (middle panel in the middle column and orange line in the right panel), peaks inside the reconnection region.
 
The total energy-momentum density tensor encompasses the ram pressure tensor, consisting of mean species velocities, and the thermal pressure tensor, consisting of thermal particle motion. 
When considering the gradients of momentum fluxes it becomes immediately evident from the previous analysis together with Ohm's law (\ref{eq:ohmslawz}) that it is only the gradients of thermal momentum fluxes that matter for supplying the reconnection electric field, regardless of how its total pressure and ram pressure analogues individually contribute.   
Here, $xz$-thermal is negligibly small (magenta line in the right panel) with $xz$-total (middle panel in the left column and dark brown line in the right panel) and $xz$-ram having similar profiles but of opposite sign.
Furthermore, $yz$-thermal (bottom panel in the middle column and dark purple line in the right panel) is similar to $yz$-ram, because $yz$-total (light brown line in the right panel) vanishes outside and is small inside the reconnection region.

This analysis shows that near the X-point the contributions of individual Ohm's law terms to the electric field have spatial distributions forming a substructure within the reconnection layer.
While just outside the reconnection region several terms contribute significantly to the electric field, inside the reconnection region only the gradients of momentum fluxes contribute to the reconnection electric field. 
Furthermore, it shows that the reconnection region captures well the inner region of the substructure that is extremely localized around the X-point where reconnection occurs and excludes the outer regions of the reconnection layer away from the X-point, thereby justifying our choice of definition of the reconnection region. 
Moreover, this analysis indicates that in the reconnection region without guide field $yz$-thermal is the dominant term that drives the reconnection electric field.

\subsection{Reconnection in the full plasmoid chain without guide field}
In a fully developed self-similar hierarchical plasmoid chain, many different X-points exist varying in their locations (e.g., in the main current sheet between plasmoids and in current sheets between merging plasmoids),  sizes, and (relativistic) velocities. 
In order to identify the reconnection mechanisms providing the reconnection electric field in all X-points of the full plasmoid chain, we use a statistical approach. 
We determine the contributions of Ohm's law terms to the reconnection electric field in the entire reconnection region of all X-points simultaneously during the entire statistical steady state.
 
\begin{figure*}[ht!]
\centering
\resizebox{\textwidth}{!}{  
\plotone{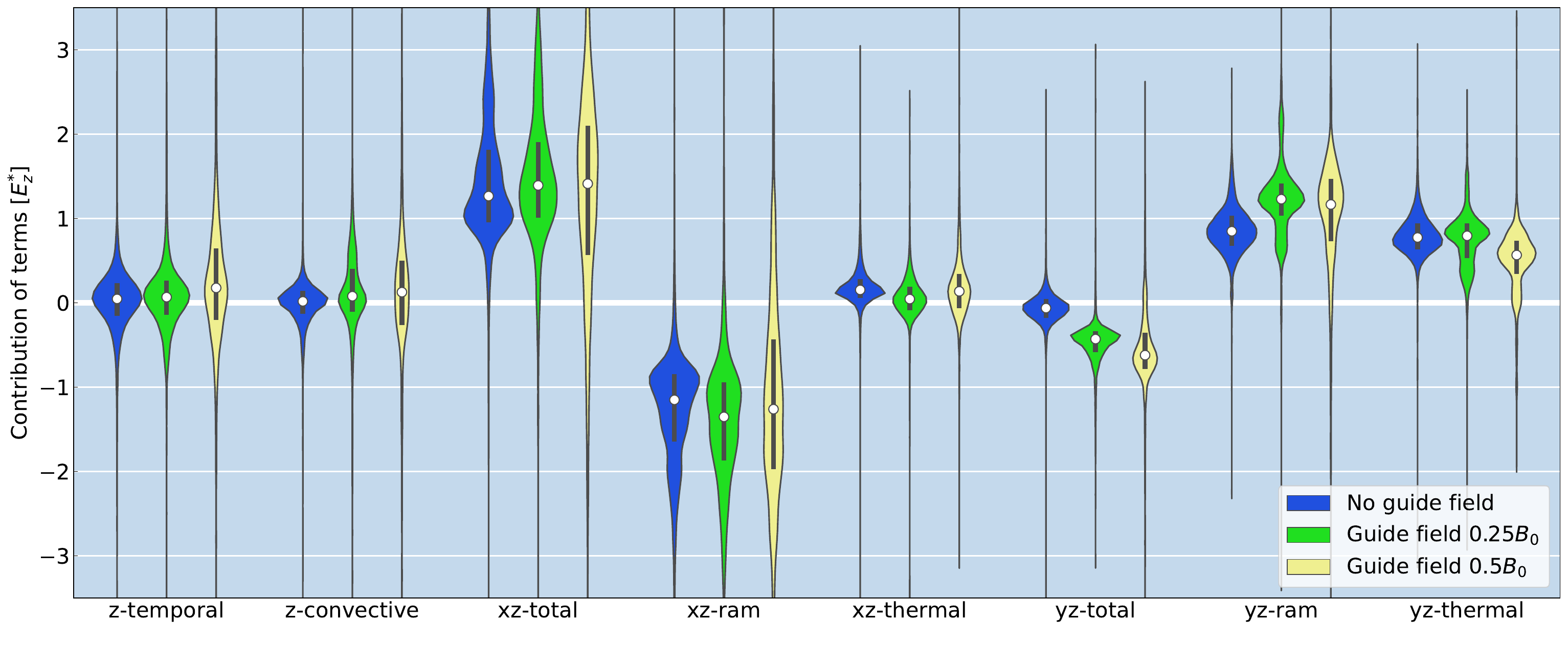}
}
\caption{
Violin plots showing the kernel density estimations (KDEs) of the distributions of Ohm’s law terms in all reconnection regions for the entire statistical steady state grouped by guide field strength for each term. 
Locally (in each cell of the reconnection region) each term is normalized to the local non-ideal electric field such that the KDEs (colored regions) show the relative contribution of the terms to the local reconnection electric field. 
Thick black bars indicate interquartile ranges (i.e.\ the middle 50\% of data points) and white dots indicate the medians.
}
\label{fig:violinplot}
\end{figure*}

Violin plots in Figure \ref{fig:violinplot} show the distributions of Ohm's law terms grouped by guide field strength for each term. Herein, the distributions are shown by kernel density estimations (KDEs) that are rotated on their side such that the distributions run over the vertical axis showing their median and spread.
Moreover, the KDEs are mirrored along the vertical center-line of each ``violin'' on which thick black bars indicate interquartile ranges (i.e.\ the middle 50\% of data points, which is a measure for the spread) and white dots indicate the medians. 
In each cell of the reconnection regions each contribution of the Ohm's law terms is normalized to the local non-ideal electric field such that these KDEs represent the relative contribution to the reconnection electric field. 

In the case of no guide field (left KDEs in blue in each grouped term in Figure \ref{fig:violinplot}) the distributions of $z$-temporal and $z$-convective are centered around zero with a relatively small spread. 
Therefore, in a statistical sense the contributions of these terms to the reconnection electric field vanish, which is expected 
based on the arguments presented in the previous analysis of a single X-point. 

The distribution of $xz$-thermal is centered near zero with a small spread making it statistically negligible, with distributions of $xz$-total and $xz$-ram that are similarly shaped but opposite in sign.
From equation (\ref{eq:ohmslawz}) this automatically means that only $yz$-thermal remains to supply the reconnection electric field.
The distribution of $yz$-thermal is similar to $yz$-ram as the contribution of $yz$-total is statistically very small.

The contributions of Ohm's law terms obtained in this statistical analysis of all X-points is in agreement with those obtained in the previous analysis of a single X-point.
We, therefore, show for collisionless reconnection in the plasmoid dominated regime without guide field that $yz$-thermal is the dominant term that drives the reconnection electric field.
Moreover, without guide field it is justified to approximate this driving term by its ram pressure analogue (since $yz$-total is negligible) similar to \citet[]{Bessho2007}, but now shown for all secondary X-points in a fully developed plasmoid chain.

This allows us to establish a useful expression for the effective resistivity.
By first reducing the full Ohm's law by neglecting all terms except $yz$-ram and then using the product rule the remaining terms are expanded into 3 subterms for each species $\alpha$ to 
$  
\partial_{y} (\gls{ndens}_{\alpha}) \langle \gls{3velcomp}_{\alpha y} \rangle \langle \gls{4velcomp}_{\alpha z} \rangle
+
\gls{ndens}_{\alpha} \partial_{y} (\langle \gls{3velcomp}_{\alpha y } \rangle ) \langle \gls{4velcomp}_{\alpha z} \rangle
+
\gls{ndens}_{\alpha} \langle \gls{3velcomp}_{\alpha y} \rangle  \partial_{y} (\langle \gls{4velcomp}_{\alpha z} \rangle)
$.
From the previous analysis of a single X-point we conclude that inside the reconnection region $\langle \gls{3velcomp}_{\alpha y} \rangle$ vanishes at the stagnation point such that  $\langle \gls{3velcomp}_{\alpha y} \rangle \ll \langle  \gls{4velcomp}_{\alpha z} \rangle $ and $\partial_{y} \langle \gls{4velcomp}_{\alpha z} \rangle \ll \partial_{y} \langle \gls{3velcomp}_{\alpha y} \rangle $.
This leads to $\gls{ndens}_{\alpha} \partial_{y} (\langle \gls{3velcomp}_{\alpha y } \rangle ) \langle \gls{4velcomp}_{\alpha z} \rangle$ being the dominant subterm which is verified by our simulations.
By simplifying Ohm's law (\ref{eq:ohmslawz}) to the form $\gls{Ecomp}_{ z}^{*} = \gls{eta}_{\textrm{eff}} \gls{3curcomp}_{ z}$ the ultimate effects of collisionless reconnection are gathered into an effective resistivity\footnote{This Ohm's law is actually only well defined in the co-moving frame, however we determined that for our case with only mildly relativistic bulk velocities in the reconnection regions, the differences arising from defining Ohm's law in the simulation domain frame versus the co-moving frame is negligible.}.
Assuming charge neutrality $\gls{ndens}_{p} = \gls{ndens}_{e}$, from symmetry $- \langle \gls{4velcomp}_{pz} \rangle =\langle \gls{4velcomp}_{ez} \rangle $ and $\langle \gls{3velcomp}_{py} \rangle =\langle \gls{3velcomp}_{ey} \rangle 
=
\langle \gls{3velcomp}_{y} \rangle
$ (all also shown from our simulations), 
and writing the dominant subterm in the form $\gls{eta}_{\textrm{eff}} \gls{3curcomp}_{z}$, with  $\gls{3curcomp}_{z} = - \gls{ndens}_{t} \gls{elemch} \langle  \gls{3velcomp}_{\alpha z} \rangle  $, an expression for the effective resistivity in a pair plasma without guide field is given by, 
\begin{eqnarray}
\gls{eta}_{\textrm{eff}} &&
= 
\frac{\gls{pmass}}{\gls{ndens}_{t} \gls{elemch}^{2}} 
\frac{\langle \gls{4velcomp}_{\alpha z} \rangle}{\langle \gls{3velcomp}_{\alpha z} \rangle}  
\frac{\partial \langle \gls{3velcomp}_{y} \rangle }{\partial y}
\label{eq:etaeff}
\end{eqnarray} 
where $\alpha$ is one of either species. The expression is
similar to \citet[]{Bessho2012}, but now also shown to be valid for all reconnection regions in a fully developed plasmoid chain.
The electric field $\gls{etaeff} \gls{3curcomp}_{z}$ (dashed lightblue line in the right panel of Figure \ref{fig:2DdomainsAndCuts}), constructed from multiplying this expression (\ref{eq:etaeff}) for the effective resistivity back with the out-of-plane current, is an adequate approximation to the non-ideal electric field  $\gls{Ecomp}_{z}^{*}$ (dotted blue line in the right panel of Figure \ref{fig:2DdomainsAndCuts}) inside the reconnection region, especially at the center-line of the current sheet (i.e. the X-point).
Towards the boundary of the reconnection region (i.e. the intersection of the shaded and non-shaded region in the right panel of Figure \ref{fig:2DdomainsAndCuts}) $yz$-thermal starts to deviate from $\gls{Ecomp}_{z}^{*}$ as other Ohm's law terms (e.g., $z$-convective) become slightly non-zero.
Furthermore, here $\gls{etaeff} \gls{3curcomp}_{z}$ slightly deviates from $yz$-thermal because neglected terms in the expansion of $yz$-ram become slightly non-zero.
Outside the reconnection region the approximation of $yz$-thermal to $\gls{Ecomp}_{z}^{*}$ breaks down and $\gls{etaeff} \gls{3curcomp}_{z}$ peaks due to stagnation of inflow velocity through $\partial_{y} \langle \gls{3velcomp}_{y} \rangle $.

Particles flowing from the upstream towards the X-point are accelerated by the reconnection electric field in the $z$-direction, after which the outflows diverge sideways.
After a finite residence time $\tau$ in the accelerating reconnection region particles are expelled into the downstream. 
The effective collisionless resistivity 
$
\gls{eta}_{\textrm{eff}} 
=
\gls{pmass}  \langle \gls{lfac}_{\alpha z} \rangle / (\gls{ndens}_{t} \gls{elemch}^{2} \tau)
$ 
emerges from this finite residence time of the particles in the reconnection region such that $\tau = | \partial_{y} \langle \gls{3velcomp}_{\alpha y} \rangle |^{-1} \sim d / \gls{3velcomp}_{in} $, with $d$ the half-width of the current sheet and $\gls{3velcomp}_{in}$ the inflow speed into the current sheet \citep{Bessho2007}. 
This is consistent with the classical collisionless inertial resistivity described by \citet{Speiser1970}.

\subsection{Reconnection in the full plasmoid chain with guide field}
Threading the system with a guide field affects the reconnection process and has been shown to slow it down.
\citet[]{Hesse2004} find that in collisionless reconnection in the presence of moderate guide fields two length scales are associated with the substructure of the reconnection layer.
On the scale of the electron skin depth the convective contribution becomes equal to the ideal contribution.
On the scale of the electron gyroradius (measured with respect to the guide field component) the thermal pressure-based contribution becomes equal to the convective contribution.
This means that stronger guide fields lead to thinner reconnection regions (i.e.\ inner substructure of the reconnection layer on the scale of the gyroradius) requiring higher numerical resolutions to be resolved, hence our choice for guide fields of moderate strength. 
  
Although for the cases with guide field (middle green and right yellow KDEs in each grouped term in Figure \ref{fig:violinplot}) the $z$-temporal and $z$-convective distributions show some changes in their shape and spread, statistically their contributions remain nearly unaffected by the guide field and are negligible.
In contrast to the findings of \citet[]{HesseZenitani2007}, our statistical results of the entire plasmoid chain do not show a significant contribution of the $z$-temporal term. 
However, we are not in the same parameter regime as we employ more moderate guide fields (in contrast to their guide field strength of $1.5 B_{0}$) and initialize particles with a lower temperature.
Furthermore, their X-point is not part of a fully developed plasmoid chain in which secondary X-points form independently of the initial conditions. Moreover, the X-point may be going through a transient phase which can potentially make them sensitive to the time-dependent term in Ohm's law. 
The $xz$-thermal contribution remains statistically negligible as distributions of $xz$-total and $xz$-ram show some changes but remain similarly shaped and opposite in sign.
The terms considered until now thus effectively remain unaffected by the guide fields and vanish, which automatically means that the only remaining term that matters for supplying the reconnection electric field regardless of the guide field strength is $yz$-thermal, as shown in Figure \ref{fig:violinplot}.
However, we find that a guide field does have a significant effect on the contributions of $yz$-total and $yz$-ram. With increasing guide field strength the distribution of $yz$-total departs significantly from zero while retaining its relatively narrow spread. 
Moreover, the distribution of $yz$-ram shifts to become (in a statistical sense) larger than the reconnection electric field (i.e., centered around a value larger than $1$ in Figure \ref{fig:violinplot}).

Here we have shown that relativistic collisionless plasmoid-mediated reconnection (up to a guide field of $0.5\gls{Bcomp}_{0}$) is predominantly driven by 
thermal pressure tensor gradients; 
more precisely in a coordinate-independent description by the gradient, perpendicular to the current sheet, 
of flux, into the direction of the main current,
of thermal momentum, perpendicular to the current
sheet. 
Furthermore, the fact that, in the presence of a guide field, $yz$-total departs statistically significantly from zero automatically means that $yz$-thermal cannot be approximated very well by $yz$-ram. 
Therefore, the expression for the effective resistivity with fluid-like quantities, initially proposed by \citet[]{Bessho2012} for a single X-point and shown above in equation (\ref{eq:etaeff}) to be valid for a full plasmoid chain of zero guide field, is not valid anymore in the case with guide field.
In order to obtain an expression with single-fluid-like quantities available in an MHD description for the case with guide field, it would be required to model the nongyrotropic thermal pressure component $\gls{prtcomp}_{\alpha yz}$, which is beyond the scope of this work \citep{Hesse2004,Most2022}.

\begin{figure}[ht!]
\centering
\resizebox{\columnwidth}{!}{ 
\plotone{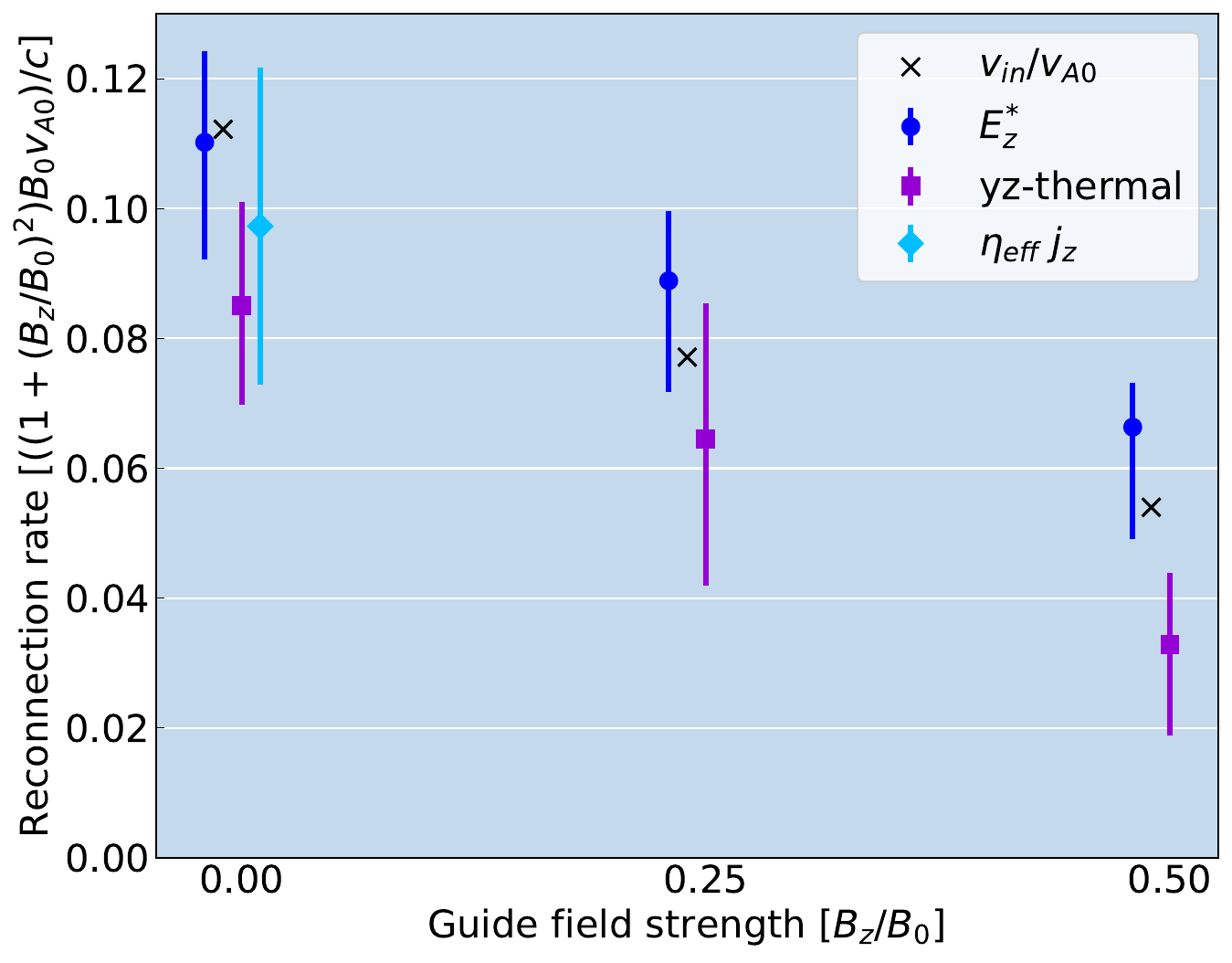}
}
\caption{
The medians and interquartile ranges of the distributions in units of $
  (   (1+\left( \gls{Bcomp}_{z} / \gls{Bcomp}_{0}\right)^{2}  )  \gls{Bcomp}_{0} \gls{valfven0}   ) /  \gls{spol}
$
of $\gls{Ecomp}_{z}^{*}$ (blue) and $yz$-thermal (dark purple) for all three guide field strengths and of $\gls{etaeff} \gls{3curcomp}_{z}$ (lightblue) for no guide field.
Stronger guide fields lead to weaker reconnection electric fields. The dominant term $yz$-thermal decreases with the reconnection electric field. 
The measured inflow velocities (black crosses), due to an E$\times$B-drift, are in close agreement with the reconnection electric fields $\gls{Ecomp}_{z}^{*}$, which directly corresponds to the reconnection rate.}
\label{fig:reconnectionrate}
\end{figure}

In Figure \ref{fig:reconnectionrate}, the medians (colored dots, squares, and diamond) and interquartile ranges (colored vertical bars) of the distributions of $\gls{Ecomp}_{z}^{*}$ (blue and dots) and $yz$-thermal (dark purple and squares) are shown for the three guide field strengths and $\gls{etaeff} \gls{3curcomp}_{z}$ (lightblue and diamond) is shown only for no guide field. These quantities are in units of
$
  (   (1+\left( \gls{Bcomp}_{z} / \gls{Bcomp}_{0}\right)^{2}  )  \gls{Bcomp}_{0} \gls{valfven0}   ) 
  /  \gls{spol}
$
comparable to a normalized reconnection rate.
For no guide field the electric field $\gls{etaeff} \gls{3curcomp}_{z}$ confirms our conclusion that the reduced $yz$-thermal term from Ohm's law, resulting in the expression (eq. \ref{eq:etaeff}) for the effective resistivity, adequately approximates the reconnection electric field.
This figure shows that a stronger guide field weakens the reconnection electric field.
Up to a guide field strength of $0.25\gls{Bcomp}_{0}$ the dominant term $yz$-thermal adequately approximates the reconnection electric field (i.e., reconnection rate), while for a guide field of $0.5 \gls{Bcomp}_{0}$ the deviation is larger.
The reconnection rate obtained from the reconnection electric field is in agreement with the actual measured inflow velocity (black crosses).

\section{Discussion and conclusions} 
In this work we identified the mechanisms driving the reconnection electric field for relativistic plasmoid-mediated reconnection in a collisionless pair plasma using a full Ohm's law by means of a statistical analysis based on our 2D fully kinetic PIC simulations.

There are, however, certain limitations to our study.
Although reconnection studies in 2D can capture many reconnection properties such as magnetic energy dissipation and particle acceleration as shown in PIC simulations \citep{Sironi2014,Sironi2022}, in reality, reconnection occurs in 3D, where plasma turbulence may play a (dominant) role in the dissipation process and may affect a resistivity model.
The plasma magnetization in neutron star or black hole magnetospheres is typically in the highly relativistic regime ($\gls{magn} \gg 1$), in accretion disks in the trans-relativistic regime ($\gls{magn} \lesssim 1$) and in black hole jet sheaths \citep{Nathanail2020,Ripperda2020,Chashkina2021} it can be asymmetric ($\gls{magn} \gg 1$ on one side, $\gls{magn} \lesssim 1$ on the other side). 
Asymmetric relativistic reconnection with different upstream properties on either side can result in different reconnection properties that are set by the upstream region with lowest magnetization \citep{Mbarek2022}.
For the highly relativistic regime, MHD simulations cannot achieve realistic plasma magnetizations due to density floors, whereas force-free (i.e., the limit of infinite magnetization) is not suitable to describe reconnection and magnetic dissipation. PIC simulations, while capturing all the kinetic physics, cannot achieve a realistic scale separation between the astrophysical system size and skin depth.
We conduct our reconnection analysis for $\gls{magn} = 10$, giving a hierarchy of scales expected for the highly magnetized case such that the gyroradius is smaller than the electron skin depth which in turn is smaller than the global scales of, for example, the length of macroscopic current sheets.
In order to understand the resistivity model for reconnection in all these regimes it would be useful to study a range of magnetizations, from the non-relativistic limit to the near-force-free limit.
Our study focused on pair plasmas, whereas in many astrophysical applications ions may be a significant component of the plasma composition, affecting the reconnection.
Yet, electron-ion reconnection in the highly relativistic regime should behave similarly to electron-positron reconnection \citep{Guo2016,Werner2018}.
In the radiative reconnection regime, that may be applicable around neutron stars and black holes \citep{Uzdensky2016book,Beloborodov2017}, the interplay between radiation and collisionless plasma can result in extensive pair production which affects the reconnection properties \citep{Hakobyan2019} and, therewith, potentially the resistivity model. 
In our analysis we consider only X-points as the non-ideal regions, while in general other non-ideal regions may exist. However, X-points are likely to be the regions with the strongest non-ideal effects.

For the case without guide field from the dominant driving mechanism an effective resistivity expression is formulated which is spatially non-uniform and peaks around the X-points.
From this, we are able to estimate realistic values for the resistivity in an astrophysical system such as for the flaring region at the jet base in M87$^{*}$ (see Appendix \ref{app:applicationtoM87}). 
Based on \citet{Ripperda2022,Hakobyan2023} we estimate, using equation \ref{eq:etaeff}, an effective local X-point resistivity $\gls{etaeff} \sim  2.7 \cdot 10^{-1} [   \textrm{s}   ]$  which is larger than the magnetospheric upstream Spitzer resistivity $\gls{eta}_{up} \sim 1.2 \cdot 10^{-24} [   \textrm{s}   ]$ by a factor of $\gls{etaeff} / \gls{eta}_{up} \sim 2.3 \cdot 10^{23}$, implying a non-uniform nature of the resistivity. 
This non-uniform nature of the effective resistivity may qualitatively capture the properties of collisionless reconnection in an MHD description.

To conclude, the fully correct approach to incorporate the effects of collisionless reconnection into fluid-type models would be to include the Ohm's law terms self-consistently in a non-ideal or even two-fluid MHD description \citep{Most2022}.  
The approach we suggest as a first step 
is to determine the Ohm's law term contributions and model essential parts of these terms into an ad hoc effective resistivity that depends on local fluid quantities (e.g., \citealt{Ripperda2019a}) in resistive relativistic MHD \citep{DelZanna2016,Ripperda2019b}.
Based on the results of this work we propose a non-uniform effective resistivity which is negligible on global scales and becomes significant only locally in X-points to mimic the properties of relativistic collisionless reconnection in non-ideal MHD simulations.
This model can provide a viable opportunity to accurately model the reconnection properties powering high-energy emission and design physically-grounded global models for it.

\section*{Acknowledgements}
S.S.\ would like to thank Sasha Philippov, Hayk Hakobyan and Amir Levinson for their help. 
B.R.\ would additionally like to thank Amitava Bhattacharjee, Sasha Chernoglazov, and Elias Most for useful discussions. Support for this work was provided by NASA through the NASA Hubble Fellowship grant HST-HF2-51518.001-A awarded by the Space Telescope Science Institute, which is operated by the Association of Universities for Research in Astronomy, Incorporated, under NASA contract NAS5-26555.
F.B.\ acknowledges support from the FED-tWIN programme (profile Prf-2020-004, project ``ENERGY'') issued by BELSPO.
The use of the national computer facilities in this research was subsidized by NWO Domain Science under grant number 2021.001. 
The computational resources and services used in this work were partially provided by 
facilities supported by the Scientific Computing Core at the Flatiron Institute, a division of the Simons Foundation; 
by the VSC (Flemish Supercomputer Center), funded by the Research Foundation Flanders (FWO) and the Flemish Government – department EWI. 
R.K.\ acknowledges funding from the European Research Council (ERC) under the European Union’s Horizon 2020 research and innovation programme (grant agreement No.\ 833251 PROMINENT ERC-ADG 2018), internal funds KU Leuven project C14/19/089 TRACESpace and the Research Foundation Flanders (FWO) project G0B4521N on flares.  
L.S.\ acknowledges support from the Cottrell Scholars Award, DoE DE-SC0021254 and NSF AST-2108201.

\clearpage
\bibliography{references-jabref,more-ref}{}

\begin{thebibliography}{}
\expandafter\ifx\csname natexlab\endcsname\relax\def\natexlab#1{#1}\fi
\providecommand{\url}[1]{\href{#1}{#1}}
\providecommand{\dodoi}[1]{doi:~\href{http://doi.org/#1}{\nolinkurl{#1}}}
\providecommand{\doeprint}[1]{\href{http://ascl.net/#1}{\nolinkurl{http://ascl.net/#1}}}
\providecommand{\doarXiv}[1]{\href{https://arxiv.org/abs/#1}{\nolinkurl{https://arxiv.org/abs/#1}}}

\bibitem[{Ball {et~al.}(2016)Ball, Özel, Psaltis, \& kwan Chan}]{Ball2016}
Ball, D., Özel, F., Psaltis, D., \& kwan Chan, C. 2016, The Astrophysical Journal, 826, 77, \dodoi{10.3847/0004-637X/826/1/77}

\bibitem[{Beloborodov(2017)}]{Beloborodov2017}
Beloborodov, A.~M. 2017, The Astrophysical Journal, 850, 141, \dodoi{10.3847/1538-4357/aa8f4f}

\bibitem[{Bessho \& Bhattacharjee(2007)}]{Bessho2007}
Bessho, N., \& Bhattacharjee, A. 2007, Physics of Plasmas, 14, 056503, \dodoi{10.1063/1.2714020}

\bibitem[{{Bessho} \& {Bhattacharjee}(2012)}]{Bessho2012}
{Bessho}, N., \& {Bhattacharjee}, A. 2012, \apj, 750, 129, \dodoi{10.1088/0004-637X/750/2/129}

\bibitem[{Bransgrove {et~al.}(2021)Bransgrove, Ripperda, \& Philippov}]{Bransgrove2021}
Bransgrove, A., Ripperda, B., \& Philippov, A. 2021, Physical Review Letters, 127, \dodoi{10.1103/physrevlett.127.055101}

\bibitem[{Broderick \& Loeb(2006)}]{Broderick2006}
Broderick, A.~E., \& Loeb, A. 2006, Monthly Notices of the Royal Astronomical Society, 367, 905, \dodoi{10.1111/j.1365-2966.2006.10152.x}

\bibitem[{Cerutti {et~al.}(2012)Cerutti, Uzdensky, \& Begelman}]{Cerutti2012}
Cerutti, B., Uzdensky, D.~A., \& Begelman, M.~C. 2012, The Astrophysical Journal, 746, 148, \dodoi{10.1088/0004-637x/746/2/148}

\bibitem[{Chashkina {et~al.}(2021)Chashkina, Bromberg, \& Levinson}]{Chashkina2021}
Chashkina, A., Bromberg, O., \& Levinson, A. 2021, Monthly Notices of the Royal Astronomical Society, 508, 1241, \dodoi{10.1093/mnras/stab2513}

\bibitem[{Comisso \& Bhattacharjee(2016)}]{ComissoBhattacharjee2016}
Comisso, L., \& Bhattacharjee, A. 2016, Journal of Plasma Physics, 82, 595820601, \dodoi{10.1017/S002237781600101X}

\bibitem[{Del~Zanna {et~al.}(2016)Del~Zanna, Papini, Landi, Bugli, \& Bucciantini}]{DelZanna2016}
Del~Zanna, L., Papini, E., Landi, S., Bugli, M., \& Bucciantini, N. 2016, Monthly Notices of the Royal Astronomical Society, 460, 3753, \dodoi{10.1093/mnras/stw1242}

\bibitem[{Dexter {et~al.}(2020)Dexter, Tchekhovskoy, Jim{\'{e} }nez-Rosales, Ressler, Bauböck, Dallilar, de~Zeeuw, Eisenhauer, von Fellenberg, Gao, Genzel, Gillessen, Habibi, Ott, Stadler, Straub, \& Widmann}]{Dexter2020}
Dexter, J., Tchekhovskoy, A., Jim{\'{e} }nez-Rosales, A., {et~al.} 2020, Monthly Notices of the Royal Astronomical Society, 497, 4999, \dodoi{10.1093/mnras/staa2288}

\bibitem[{Giannios(2013)}]{Giannios2013}
Giannios, D. 2013, Monthly Notices of the Royal Astronomical Society, 431, 355, \dodoi{10.1093/mnras/stt167}

\bibitem[{Guo {et~al.}(2016)Guo, Li, Li, Daughton, Zhang, Lloyd-Ronning, Liu, Zhang, \& Deng}]{Guo2016}
Guo, F., Li, X., Li, H., {et~al.} 2016, The Astrophysical Journal Letters, 818, L9, \dodoi{10.3847/2041-8205/818/1/L9}

\bibitem[{Hakobyan {et~al.}(2019)Hakobyan, Philippov, \& Spitkovsky}]{Hakobyan2019}
Hakobyan, H., Philippov, A., \& Spitkovsky, A. 2019, The Astrophysical Journal, 877, 53, \dodoi{10.3847/1538-4357/ab191b}

\bibitem[{Hakobyan {et~al.}(2023)Hakobyan, Ripperda, \& Philippov}]{Hakobyan2023}
Hakobyan, H., Ripperda, B., \& Philippov, A.~A. 2023, The Astrophysical Journal Letters, 943, L29, \dodoi{10.3847/2041-8213/acb264}

\bibitem[{Hakobyan \& Spitkovsky(2020)}]{Hakobyan2020}
Hakobyan, H., \& Spitkovsky, A. 2020, {multi-species particle-in-cell plasma code}, \url{https://ntoles.github.io/tristan-wiki/}

\bibitem[{{Hesse} {et~al.}(2004){Hesse}, {Kuznetsova}, \& {Birn}}]{Hesse2004}
{Hesse}, M., {Kuznetsova}, M., \& {Birn}, J. 2004, Physics of Plasmas, 11, 5387, \dodoi{10.1063/1.1795991}

\bibitem[{Hesse \& Zenitani(2007)}]{HesseZenitani2007}
Hesse, M., \& Zenitani, S. 2007, Physics of Plasmas, 14, 112102, \dodoi{10.1063/1.2801482}

\bibitem[{Hu \& Beloborodov(2021)}]{Hu2021}
Hu, R., \& Beloborodov, A.~M. 2021, Axisymmetric pulsar magnetosphere revisited,  arXiv, \dodoi{10.48550/ARXIV.2109.03935}

\bibitem[{Keppens {et~al.}(2013)Keppens, Porth, Galsgaard, Frederiksen, Restante, Lapenta, \& Parnell}]{Keppens2013}
Keppens, R., Porth, O., Galsgaard, K., {et~al.} 2013, Physics of Plasmas, 20, 092109, \dodoi{10.1063/1.4820946}

\bibitem[{Lyubarsky(2018)}]{Lyubarsky2018}
Lyubarsky, Y. 2018, Monthly Notices of the Royal Astronomical Society, 483, 1731, \dodoi{10.1093/mnras/sty3233}

\bibitem[{Lyubarsky(2020)}]{Lyubarsky2020}
---. 2020, The Astrophysical Journal, 897, 1, \dodoi{10.3847/1538-4357/ab97b5}

\bibitem[{Lyubarsky(2012)}]{Lyubarsky2012}
Lyubarsky, Y.~E. 2012, Monthly Notices of the Royal Astronomical Society, 427, 1497, \dodoi{10.1111/j.1365-2966.2012.22097.x}

\bibitem[{{Lyutikov} \& {McKinney}(2011)}]{Lyutikov2011}
{Lyutikov}, M., \& {McKinney}, J.~C. 2011, \prd, 84, 084019, \dodoi{10.1103/PhysRevD.84.084019}

\bibitem[{Lyutikov \& Popov(2020)}]{Lyutikov2020}
Lyutikov, M., \& Popov, S. 2020, Fast Radio Bursts from reconnection events in magnetar magnetospheres,  arXiv, \dodoi{10.48550/ARXIV.2005.05093}

\bibitem[{Mahlmann {et~al.}(2022)Mahlmann, Philippov, Levinson, Spitkovsky, \& Hakobyan}]{Mahlmann2022}
Mahlmann, J.~F., Philippov, A.~A., Levinson, A., Spitkovsky, A., \& Hakobyan, H. 2022, The Astrophysical Journal Letters, 932, L20, \dodoi{10.3847/2041-8213/ac7156}

\bibitem[{Mbarek {et~al.}(2022)Mbarek, Haggerty, Sironi, Shay, \& Caprioli}]{Mbarek2022}
Mbarek, R., Haggerty, C., Sironi, L., Shay, M., \& Caprioli, D. 2022, Phys. Rev. Lett., 128, 145101, \dodoi{10.1103/PhysRevLett.128.145101}

\bibitem[{Most {et~al.}(2022)Most, Noronha, \& Philippov}]{Most2022}
Most, E.~R., Noronha, J., \& Philippov, A.~A. 2022, Monthly Notices of the Royal Astronomical Society, \dodoi{10.1093/mnras/stac1435}

\bibitem[{Most \& Philippov(2020)}]{Most2020}
Most, E.~R., \& Philippov, A.~A. 2020, The Astrophysical Journal, 893, L6, \dodoi{10.3847/2041-8213/ab8196}

\bibitem[{Most \& Philippov(2022)}]{Most2022b}
---. 2022, Reconnection-powered fast radio transients from coalescing neutron star binaries,  arXiv, \dodoi{10.48550/ARXIV.2207.14435}

\bibitem[{Nathanail {et~al.}(2020)Nathanail, Fromm, Porth, Olivares, Younsi, Mizuno, \& Rezzolla}]{Nathanail2020}
Nathanail, A., Fromm, C.~M., Porth, O., {et~al.} 2020, Monthly Notices of the Royal Astronomical Society, 495, 1549, \dodoi{10.1093/mnras/staa1165}

\bibitem[{Nathanail {et~al.}(2022)Nathanail, Mpisketzis, Porth, Fromm, \& Rezzolla}]{Nathanail2022}
Nathanail, A., Mpisketzis, V., Porth, O., Fromm, C.~M., \& Rezzolla, L. 2022, Monthly Notices of the Royal Astronomical Society, 513, 4267, \dodoi{10.1093/mnras/stac1118}

\bibitem[{Parfrey {et~al.}(2012)Parfrey, Beloborodov, \& Hui}]{Parfrey2012}
Parfrey, K., Beloborodov, A.~M., \& Hui, L. 2012, The Astrophysical Journal, 754, L12, \dodoi{10.1088/2041-8205/754/1/l12}

\bibitem[{Petropoulou {et~al.}(2016)Petropoulou, Giannios, \& Sironi}]{Petropoulou2016}
Petropoulou, M., Giannios, D., \& Sironi, L. 2016, Monthly Notices of the Royal Astronomical Society, 462, 3325, \dodoi{10.1093/mnras/stw1832}

\bibitem[{Philippov {et~al.}(2019)Philippov, Uzdensky, Spitkovsky, \& Cerutti}]{Philippov2019}
Philippov, A., Uzdensky, D.~A., Spitkovsky, A., \& Cerutti, B. 2019, The Astrophysical Journal, 876, L6, \dodoi{10.3847/2041-8213/ab1590}

\bibitem[{Philippov \& Spitkovsky(2018)}]{Philippov2018}
Philippov, A.~A., \& Spitkovsky, A. 2018, The Astrophysical Journal, 855, 94, \dodoi{10.3847/1538-4357/aaabbc}

\bibitem[{Porth {et~al.}(2021)Porth, Mizuno, Younsi, \& Fromm}]{Porth2021}
Porth, O., Mizuno, Y., Younsi, Z., \& Fromm, C.~M. 2021, Monthly Notices of the Royal Astronomical Society, 502, 2023, \dodoi{10.1093/mnras/stab163}

\bibitem[{Ripperda {et~al.}(2020)Ripperda, Bacchini, \& Philippov}]{Ripperda2020}
Ripperda, B., Bacchini, F., \& Philippov, A.~A. 2020, The Astrophysical Journal, 900, 100, \dodoi{10.3847/1538-4357/ababab}

\bibitem[{Ripperda {et~al.}(2022)Ripperda, Liska, Chatterjee, Musoke, Philippov, Markoff, Tchekhovskoy, \& Younsi}]{Ripperda2022}
Ripperda, B., Liska, M., Chatterjee, K., {et~al.} 2022, The Astrophysical Journal Letters, 924, L32, \dodoi{10.3847/2041-8213/ac46a1}

\bibitem[{Ripperda {et~al.}(2019)Ripperda, Porth, Sironi, \& Keppens}]{Ripperda2019a}
Ripperda, B., Porth, O., Sironi, L., \& Keppens, R. 2019, Monthly Notices of the Royal Astronomical Society, 485, 299, \dodoi{10.1093/mnras/stz387}

\bibitem[{{Ripperda} {et~al.}(2019){Ripperda}, {Bacchini}, {Porth}, {Most}, {Olivares}, {Nathanail}, {Rezzolla}, {Teunissen}, \& {Keppens}}]{Ripperda2019b}
{Ripperda}, B., {Bacchini}, F., {Porth}, O., {et~al.} 2019, \apjs, 244, 10, \dodoi{10.3847/1538-4365/ab3922}

\bibitem[{Scepi {et~al.}(2022)Scepi, Dexter, \& Begelman}]{Scepi2022}
Scepi, N., Dexter, J., \& Begelman, M.~C. 2022, Monthly Notices of the Royal Astronomical Society, 511, 3536, \dodoi{10.1093/mnras/stac337}

\bibitem[{Sironi(2022)}]{Sironi2022}
Sironi, L. 2022, Physical Review Letters, 128, \dodoi{10.1103/physrevlett.128.145102}

\bibitem[{Sironi \& Beloborodov(2020)}]{Sironi2020}
Sironi, L., \& Beloborodov, A.~M. 2020, The Astrophysical Journal, 899, 52, \dodoi{10.3847/1538-4357/aba622}

\bibitem[{Sironi \& Spitkovsky(2014)}]{Sironi2014}
Sironi, L., \& Spitkovsky, A. 2014, The Astrophysical Journal, 783, L21, \dodoi{10.1088/2041-8205/783/1/l21}

\bibitem[{Speiser(1970)}]{Speiser1970}
Speiser, T. 1970, Planetary and Space Science, 18, 613, \dodoi{https://doi.org/10.1016/0032-0633(70)90136-4}

\bibitem[{Sridhar {et~al.}(2021)Sridhar, Sironi, \& Beloborodov}]{Sridhar2021}
Sridhar, N., Sironi, L., \& Beloborodov, A.~M. 2021, Monthly Notices of the Royal Astronomical Society, 507, 5625, \dodoi{10.1093/mnras/stab2534}

\bibitem[{Sridhar {et~al.}(2022)Sridhar, Sironi, \& Beloborodov}]{Sridhar2022}
---. 2022, Comptonization by Reconnection Plasmoids in Black Hole Coronae II: Electron-Ion Plasma,  arXiv, \dodoi{10.48550/ARXIV.2203.02856}

\bibitem[{Uzdensky(2016)}]{Uzdensky2016book}
Uzdensky, D.~A. 2016, in Magnetic Reconnection (Springer International Publishing), 473--519, \dodoi{10.1007/978-3-319-26432-5_12}

\bibitem[{{Uzdensky} {et~al.}(2010){Uzdensky}, {Loureiro}, \& {Schekochihin}}]{Uzdensky2010}
{Uzdensky}, D.~A., {Loureiro}, N.~F., \& {Schekochihin}, A.~A. 2010, \prl, 105, 235002, \dodoi{10.1103/PhysRevLett.105.235002}

\bibitem[{Werner {et~al.}(2018)Werner, Uzdensky, Begelman, Cerutti, \& Nalewajko}]{Werner2018}
Werner, G., Uzdensky, D., Begelman, M., Cerutti, B., \& Nalewajko, K. 2018, MNRAS, 473, 4

\bibitem[{Younsi \& Wu(2015)}]{Younsi2015}
Younsi, Z., \& Wu, K. 2015, Monthly Notices of the Royal Astronomical Society, 454, 3283, \dodoi{10.1093/mnras/stv2203}

\end{thebibliography}
\bibliographystyle{aasjournal}

\appendix

\section{Estimation of resistivities in flaring region of M87$^{*}$} 
\label{app:applicationtoM87}
Reconnection as a flare mechanism for active galactic nuclei (AGN) has been proposed by \citet[]{Broderick2006,Younsi2015, Ball2016,Dexter2020,Porth2021,Scepi2022} and shown to occur in 2D by \cite{Nathanail2020,Ripperda2020,Chashkina2021} and in 3D by \cite{Ripperda2022,Nathanail2022}. 
In the magnetically arrested disk model, the most energetic flares originate close to the event horizon \citep{Ripperda2022} from reconnection sites, without guide field, fed by highly magnetized pair plasma from the jet. 
As an example for AGN flares, we use the well-constrained conditions of M87$^{*}$ to illustrate the non-uniformity between the global upstream resistivity and the local effective X-point resistivity for the flaring region at the jet base.

During the flaring state, the synchrotron-produced photons from reconnection accelerated particles in the current sheet produce secondary pairs through photon-photon collisions in the upstream.
At the time of birth these secondary pairs are estimated to have a characteristic Lorentz factor $\gls{lfac}_{\textrm{sec}} \approx 200$ \citep[]{Hakobyan2023} setting the characteristic temperature dictated by the synchrotron burnoff limit as $\gls{boltzconst} \gls{temp} / ( \gls{pmass} \gls{spol}^{2} ) = \gls{lfac}_{\textrm{sec}} / 3 \approx 200 / 3$.
This gives an upstream Spitzer resistivity due to Coulomb collisions of pairs of
$
\gls{eta}_{up}
=
(\gls{boltzconst} \gls{temp} / ( \gls{pmass} \gls{spol}^{2} ))^{-3/2}
4 \sqrt{2 \pi} \gls{elemch}^{2} \mathrm{ln}(\Lambda)/(3 \gls{pmass} \gls{spol}^{3} )
\sim
1.2 \cdot 10^{-24} [\textrm{s}  ]
$ with Coulomb logarithm $\mathrm{ln}(\Lambda) = 21$. 

Fiducial parameters for M87$^{*}$ are an upstream magnetic field strength $\gls{Bcomp}_{up} = 10^{2} [G]$ at the event horizon, a central mass $\gls{mass} = 6 \cdot 10^{9} [\gls{masssun}]$ and a multiplicity $\gls{multipl} = 10^{8}$ for the pair production during the flaring state \citep[]{Ripperda2022}. 
Using the Goldreich-Julian density leads to a particle density 
$
\gls{ndens} 
\sim
1.9 \cdot 10^{3} (\gls{multipl} / 10^{8})(\gls{mass}/ (6 \cdot 10^{9} \gls{masssun}))^{-1}(\gls{Bcomp}_{up}/10^{2} [\textrm{G}] ) [\textrm{cm}^{-3}]
$
giving a plasma skin depth 
$
\gls{skind} 
 =
 \gls{spol} / \gls{pfreq}
= 
(\gls{pmass}_{e} \gls{spol}^{2} / ( 4 \pi \gls{ndens} \gls{elemch}^{2} ))^{1/2}
\sim 
1.2 \cdot 10^{4} (\gls{multipl} / 10^{8})^{-1/2} (\gls{mass}/ (6 \cdot 10^{9} \gls{masssun}))^{1/2}(\gls{Bcomp}_{up}/10^{2} [\textrm{G}] )^{-1/2} [\textrm{cm} ]
$ 
and an upstream plasma magnetization 
$\gls{magn}_{up}
= 
\gls{Bcomp}^{2} / (4 \pi \gls{ndens} \gls{pmass} \gls{spol}^{2}) 
\sim 
5.2 \cdot 10^{5} (\gls{multipl}/ 10^{8} )^{-1} (\gls{mass}/ (6 \cdot 10^{9} \gls{masssun})) (\gls{Bcomp}_{up}/10^{2} [\textrm{G}] ) $, during the flaring state.

To estimate the X-point resistivity, we further inspect Equation (\ref{eq:etaeff}) which  can be recast into $\eta_{\rm eff}=4\pi/c^2\, \lambda_{\rm p} \langle \gls{lfac}_{\alpha z} \rangle v_{\rm in} $ where $\langle \gls{lfac}_{\alpha z} \rangle :=\langle \gls{4velcomp}_{\alpha z} \rangle / \langle \gls{3velcomp}_{\alpha z} \rangle $.  The latter factor comes from the relativistically increased inertia of the accelerated particles in the X-points and is not present in the non-relativistic treatment \citep[see also][]{Bessho2012}.   
Here we have also used that the inflow velocity $ v_{in}\sim \langle \gls{3velcomp}_{\alpha y} \rangle  \sim 0.1 \gls{spol} $ changes over the (half-)width of the current sheet on the scale of the skin depth giving $\partial_{y} \langle \gls{3velcomp}_{\alpha y} \rangle \sim \gls{3velcomp}_{in} / \gls{skind}$.  
The upstream plasma magnetization sets the available magnetic energy per particle and thus $\langle \gls{lfac}_{\alpha z} \rangle \sim \gls{magn}_{up}$ \citep{Sironi2014} such that for the fiducial M87$^{*}$ parameters we have $\gls{etaeff} \sim 4 \pi \gls{magn}_{up} \gls{skind}  0.1 / \gls{spol}  =  2.7 \cdot 10^{-1} [ \textrm{s}]$, which is larger than the upstream resistivity by a factor of $\gls{etaeff} / \gls{eta}_{up} \sim 2.3 \cdot 10^{23}$.

\end{document}